\documentclass{mn2e}

\title{Diffusion coefficient of a passive contaminant in a local MHD model of a turbulent accretion disc}

\author[]{Augusto Carballido$^{1,2}$,\thanks{E-mail: augusto@ast.cam.ac.uk}
James M. Stone$^{2}$,\thanks{E-mail: jstone@astro.princeton.edu}
and James E. Pringle$^{1}$\thanks{E-mail: jep@ast.cam.ac.uk} \\
$\rm ^1$ Institute of Astronomy, University of Cambridge, Madingley Road,
Cambridge CB3 0HA, UK \\
$\rm ^2$ Department of Astrophysical Sciences, Princeton University, Princeton
NJ 08544, USA \\}

\begin{document}
\maketitle
\begin{abstract}
We calculate the radial diffusion coefficient for a passive contaminant in
an accretion disc which is turbulent due to the action of the magnetorotational
instability.  Numerical MHD simulations are used to follow the evolution
of a local patch of the disc using the shearing box formalism.  A separate
continuity equation for the mass fraction of contaminant is integrated
along with the MHD sytem, and radial profiles of this fraction are
obtained as a function of time.  Solutions of a linear diffusion equation
are fitted to the numerical measured profiles of the contaminant, treating
the diffusion coefficient $D$ as the fitting parameter.  At early times,
the value of $D$ is found to vary, however once the contaminant is spread
over scales comparable to the box size, it saturates at a steady value.
The ratio of $D$ to the transport coefficient of angular momentum due to
shear stress is small.  If $D$ can be used as a proxy for the turbulent
magnetic diffusivity, the effective magnetic Prandtl number $P_{\rm eff}=\nu / D$
(where $\nu$
is the coefficient of "effective viscosity" associated with shear stress)
would be large.

\end{abstract}


\section{Introduction}

In order to understand the morphology and chemical composition of
protoplanetary accretion discs, the radial mixing of vapour and solid
material in accretion flows needs to be characterised (Morfill 1983).
The distribution of dust across a turbulent disc, as a consequence of
both advective and diffusive modes, will determine the mechanism which
is most favourable for the assemblage of particles into planetesimals,
meteoritic bodies and eventually planets.  For example, there is evidence
for radial upstream movement of nebular gas from the coexistence of
calcium-aluminum inclusions and chondrules in meteorites (Boss 1998).
Studies by Takeuchi \& Lin (2002) and by Keller \& Gail (2004) indicate
that, for vertically isothermal {\em laminar} discs, there may exist an
equatorial outflow capable of transporting material from the hot, inner
regions to the cold, outer regions, thereby affecting the composition of
bodies located in the latter zone. The material is assumed to be well
coupled to the gas, which is true for particles whose size is smaller
than the mean free path of the gas molecules.  However, this picture
may be changed if angular momentum transport in the disc is mediated
by magnetohydrodynamical turbulence driven by the magneto-rotational
instability (MRI) (Balbus \& Hawley 1991).

\indent 
The transport of a passive scalar has been used as a criterion to study
diffusive processes in turbulent flows (e.g. Brandenburg, K\"{a}pyl\"{a}
\& Mohammed 2004 and references therein). Research on the behaviour of 
passive tracers in turbulence has shown that there is a strong 
variability of dissipation and mixing rates of both scalar and velocity fields, due to small scale intermittency. It has also been shown that the central 
assumption of Kolmogorov scaling, that at small scales the scalar and 
velocity fields are isotropic in the limit of infinite Reynolds and Pecl\'{e}t
numbers, breaks down (Warhaft 2000; Sreenivasan 1991). Although a fascinating subject, in this work we will not address many important issues that belong to the vast field of passive scalars.

\indent Clarke \& Pringle (1988)
analysed the evolution of the concentration of a trace contaminant
in a steady, Keplerian disc. They point out a subtlety related to the
identification of a diffusion coefficient with a viscosity for turbulent
accretion. Despite the fact that both are associated with velocity
fluctuations in the turbulence, their effects are distinct: diffusion
tends to produce a uniform mixing of material, whereas viscosity does
not yield a uniform distribution of angular momentum. They found that the
contaminant behaviour depends on the ratio of the diffusion coefficient
to viscosity; only when this ratio is near or greater than unity does diffusion
take place appreciably.  Heyvaerts et al. (1996) have shown that in an
accretion disc in which the effective viscosity is due to turbulence generated
by the Balbus-Hawley instability, the associated magnetic Reynolds number
is small, whereas the magnetic Prandtl number $P_{m}= \nu/\eta$
(where $\eta$ and $\nu$ are the magnetic resistivity and the effective
viscosity, respectively) is not necessarily so.

\indent In this context, we wish to investigate the role that turbulent
diffusion plays locally in the transport of a passive tracer, within
a small portion of a weakly magnetised accretion disc in which the MRI
operates.  We use numerical MHD simulations which adopt the so-called
``shearing box'' model, discussed in detail by Hawley, Gammie \& Balbus
(1995, hereafter HGB), a useful scheme for the study of local processes
in accretion flows.  We extend the numerical scheme to explicitely follow
the time evolution of a passive contaminant in the flow to determine the
rate of turbulent mixing.  We use the ideal MHD approximation to simplify
the problem; this requires the magnetic field be well coupled to the gas.
Although the structure of protoplanetary discs is not well known, it
is likely that far from the central star, the central regions of the
disc are dense and cool enough that non-ideal MHD effects such as Ohmic
dissipation, ambipolar diffusion, and the Hall effect will significantly
alter the MHD (Blaes \& Balbus 1994; Wardle 1999; Sano et al. 2000;
Stone et al. 2000; Balbus \&
Terquem 2001; Salmeron \& Wardle 2003).  Recent studies of
the MRI in non-ideal MHD indicate turbulence and transport my be strongly
suppressed if the magnetic Reynolds number is close to one (Hawley \&
Stone 1998; Sano \& Miyama 1999; Fleming, Stone, \& Hawley 2000; Sano
\& Stone 2002).  It is possible that even if the MRI is suppressed
in the central regions of the disc, the surface layers may be ionized
by non-thermal effects and remain turbulent (Gammie 1996; Glassgold et al. 1997;
Fleming \&
Stone 2003).  Of course, by adopting the ideal MHD approximation, this
study cannot be applied to regions of the disc where the ionisation
fraction is so low that it is stable to the MRI: the turbulent transport
of contaminants in these regions will be essentially zero.  In regions
where the MRI is suppressed but not eliminated, the saturation amplitude
of the turbulence will be reduced, thereby reducing both the coefficient
of effective viscosity $\nu$ and the turbulent coefficient $D$.  Thus, the
ratio of the two (the effective magnetic Prandtl number $P_{\rm eff}$) may be
independent of the level of saturation of the instability.
Studying
diffusion in realistic, global models of protoplanetery discs including
non-ideal MHD and the vertical and radial dependence of the ionisation
fraction is a challenging problem and beyond the scope of this study.

\indent Previous work has investigated diffusive processes in
different magnetohydrodynamic contexts. Yousef et al. (2003) performed
forced turbulence simulations in which they obtain a value of the
turbulent magnetic Prandtl number of about unity, regardless of the value
of the microscopic magnetic Prandtl number. Brandenburg et al. (2004)
study nonlocal aspects of turbulent transport by adding a second-order
time derivative to the equation for the concentration of a passive
contaminant, namely Fick's law, which states the proportionality of
the flux of passive scalar to the negative concentration gradient. This
extra term results in a damped wave equation for the concentration.
Our investigation is similar, except turbulence in our models is driven
by a linear instability in the background flow, rather than by an
assumed forcing.  Moreover, the orbital dynamics in an accretion flow
strongly affect the velocity and magnetic field correlations in the turbulence,
which may directly affect the turbulent diffusion.

\indent  This paper is organised as follows.  In Section 2 the numerical
method employed is described, including the method by which the turbulent
diffusion coefficient is calculated.  Results of a variety of simulations
are described in Section 3, while a discussion and conclusions are
presented in Section 4.

\section{METHOD}
\subsection{Numerical Method}
We use a three-dimensional version of the ZEUS code (Stone \&
Norman 1992a;b) in which the shearing box model of HGB is implemented.
Briefly, a coordinate system is centered at a fiducial radius $r_{0}$
in a small patch of a disc, which has linear dimensions much less
than $r_{0}$.  This new system corrotates at an angular velocity
$\Omega_{0}=\Omega(r_{0})$. The $x,y,z$ coordinates correspond to
cilindrical $r,\varphi,z$ coordinates, respectively. The resulting MHD
equations then incorporate the effects of Coriolis forces and tidal
stresses (see below). The numerical implementation of the shearing
box involves boundary conditions such that the computational domain is
strictly periodic in the $y,z$ directions at all times, whereas in the $x$
(radial) direction it is {\em quasiperiodic}: when a fluid element exits
at one radial boundary, it reappears at the opposite radial boundary
with a position and a velocity given by the shearing motion.

The passive contaminant is introduced in the code as a fraction of the
gas density. In addition to the MHD equations for the shearing box,

\begin{equation} 
\frac{\partial \rho}{\partial t} + \nabla \cdot (\rho \textbf{v})=0
\label{eq:mhd1}
\end{equation}

\begin{equation}
\frac{\partial \textbf{v}}{\partial t} + \textbf{v} \cdot \nabla \textbf{v}=-\frac{1}{\rho} \nabla \left( P + \frac{B^2}{8\pi}\right) + \frac{(\textbf{B} \cdot \nabla) \textbf{B}}{4\pi \rho} -2\Omega \times \textbf{v} + 3\Omega^2x \hat{x}
 \label{eq:mhd2}
\end{equation}

\begin{equation}
\frac{\partial \textbf{B}}{\partial t}=\nabla \times (\textbf{v}\times \textbf{B})
 \label{eq:mhd3}
\end{equation}

\begin{equation} 
\frac{\partial \rho \epsilon}{\partial t} + \nabla \cdot (\rho \epsilon \textbf{v}) + P \nabla \cdot \textbf {v}=0
\label{eq:mhd4}
\end{equation}

\begin{equation} 
P=\rho \epsilon (\gamma-1)
\label{eq:mhd5}
\end{equation}

\noindent where the symbols have their usual meanings, the code also
solves a continuity equation for the contaminant fraction:

\begin{equation} 
\frac{\partial f \rho}{\partial t} + \nabla \cdot (f \rho \textbf{v})=0
\label{eq:contfrac}
\end{equation}

\noindent where $f$ is a number between 0 and 1.

As in HGB, we use the standard values $P=P_{0}=10^{-6}$ (initial
pressure), $H=1$ (box height), and $\Omega=10^{-3}$.
Our runs adopt the same parameter values as those
of model Z4 of HGB (in particular, we use a uniform strength
vertical field with plasma parameter $\beta=400$),
except that we use a resolution of $88 \times 184 \times 88$ grid zones,
almost three times greater in each dimension than that used in Z4.
The box size is $1\times 2\pi \times 1$.

\subsection{Simulations}
HGB showed that turbulence develops after a few orbits in the evolution
of the shearing box (one orbit=$2\pi/\Omega$). We first run the code
for 16 orbits, to ensure that turbulence has set in and is able to stir
the contaminant efficiently. At $t=16$ orbits the code is restarted and
the contaminant is introduced; we set $f=1$ for $-0.15 \leq x \leq 0.15$
and for all $y$ and $z$, forming a slab inside the box.  The system is
then allowed to evolve for several more orbits, and the rate of radial 
spreading of the slab is used to determine the turbulent diffusion
coefficient.

We expect statistical fluctuations in the evolution of $f$ to be large
given the chaotic nature of turbulence driven by the MRI (Winters, Balbus,
\& Hawley 2003).  Thus, to improve our measurement we have performed
22 runs beginning at different times, reintroducing the contaminant in
its original distribution each time a run is restarted, and evolving the
flow for different periods.  We shall group these simulations according
to the length they were run, labeling each group with different letters,
and each run within a group by different numbers.  Run A1 was evolved
for one orbit beginning at orbit 16.  Runs B1 through B8 were evolved
sequentially for 0.15 orbits each, starting at orbit 17 (so that run B1
provides a measurement of how the contaminant mixes in the flow during
the evolution from 17.00 to 17.15 orbits; run B2 from 17.15 to 17.30
orbits, etc.).  Runs C1 through C13 were evolved sequentially for 3
orbits each, starting at 18.2 orbits (so that run C1 is evolved from
18.2 to 21.2 orbits, etc.).  In this way we are able to measure the
turbulent mixing in the flow over 41.2 orbits of evolution.

From each run we obtain radial profiles of the azimuthally and vertically
averaged contaminant fraction $f$ at intervals of 0.05 orbits.  For each
one of these snapshots, we average the corresponding radial curves
over the number of runs performed, and hence obtain a single curve at
each time.  For example, to obtain the curve that corresponds to the
radial profile of $f$ at 0.15 orbits, we can average over all 22 runs
(since all were run for at least that long).  However, for the curve
that corresponds to the radial profile of $f$ at 1.5 orbits, we average
over the 13 runs in the group labelled C.

\subsection{Calculation of diffusion coefficient}
In order to calculate a value of the diffusion coefficient associated
with the turbulent flow, we have chosen to solve a diffusion equation
for the contaminant fraction in the radial direction with the same
initial condition as that introduced in the code:

\begin{equation} \label{eq:diffeq}
\frac{\partial f}{\partial t}=D \frac{\partial^2 f}{\partial x^2},
\end{equation}

\begin{displaymath}
f(x,0)=\left\{ \begin{array}{ll}
1 & \textrm{if $|x|\leq 0.15$}\\
0 & \textrm{if $|x| > 0.15$}
 \end{array} \right.
\end{displaymath}

The solution of this initial value problem is

\begin{equation} \label{eq:sol}
f(x,t)=\frac{1}{2}\biggl[ \textrm{erf} \Big( \frac{x+0.15}{2 \sqrt{D t}} \Big) - \textrm{erf} \Big( \frac{x-0.15}{2 \sqrt{D t}} \Big) \biggr]
\end{equation}

We then fit this solution to the curves obtained from the simulations,
treating the diffusion coefficient $D$ as the fitting parameter.  To test
whether $D$ is constant in time, the
fitting is performed independently for each time in the evolution
of the contaminant.

\indent It is worth pointing out that we are using equation (7) as a first 
approximation, and it may not describe the diffusion process of the
contaminant completely. 

\section{RESULTS}

Figure 1 shows the evolution of the $f=0.5$ isosurface of the contaminant
in Run A1 at $t=16$, 16.05 and 16.1 orbits (that is at the start of the run,
and 0.05 and 0.1 orbits later).  At early times,
wrinkling of the surfaces  on small scales is evident, as radial advective
transport of the contaminant by turbulent eddies occurs.  At later times,
larger scale distortions appear, as the contaminant is advected on larger
scales by larger eddies.

\begin{figure*}
\begin{picture}(450,317)
\put(0,0){\includegraphics{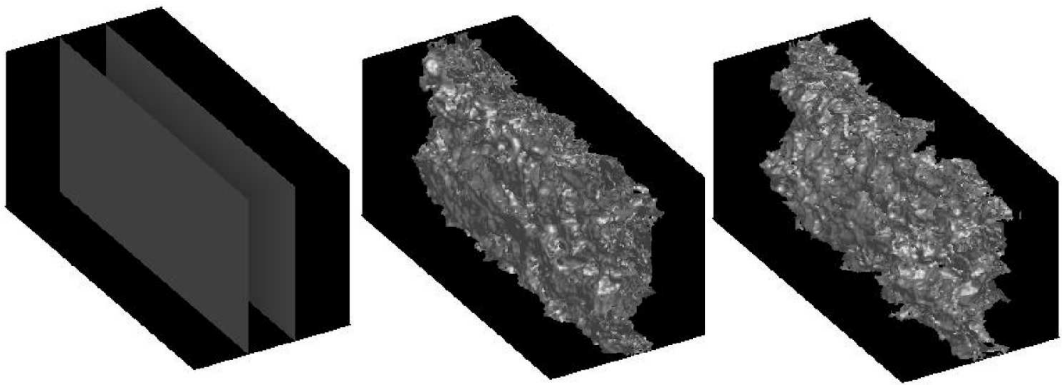}}
\end{picture}
\caption{Evolution of the $f=0.5$ isosurface of contaminant in Run A,
starting at $t=16$ orbits (left panel) when the contaminant is introduced
into the saturated turbulence, $t=16.05$ orbits (center panel), and $t=16.1$
orbits (right panel).  } \end{figure*}

Figure 2 plots the azimuthally and vertically averaged radial profile
of the contaminant from $t=0.05$ orbits to $t=2.75$ orbits after the
contaminant is introduced.  The averages are taken from all the data
available at each time.  Over this timescale, steady diffusion of the
contaminant towards a uniform profile (consistent with the radially
periodic boundary conditions) is evident.

\begin{figure*}
\begin{picture}(505,361)
\put(0,0){\includegraphics{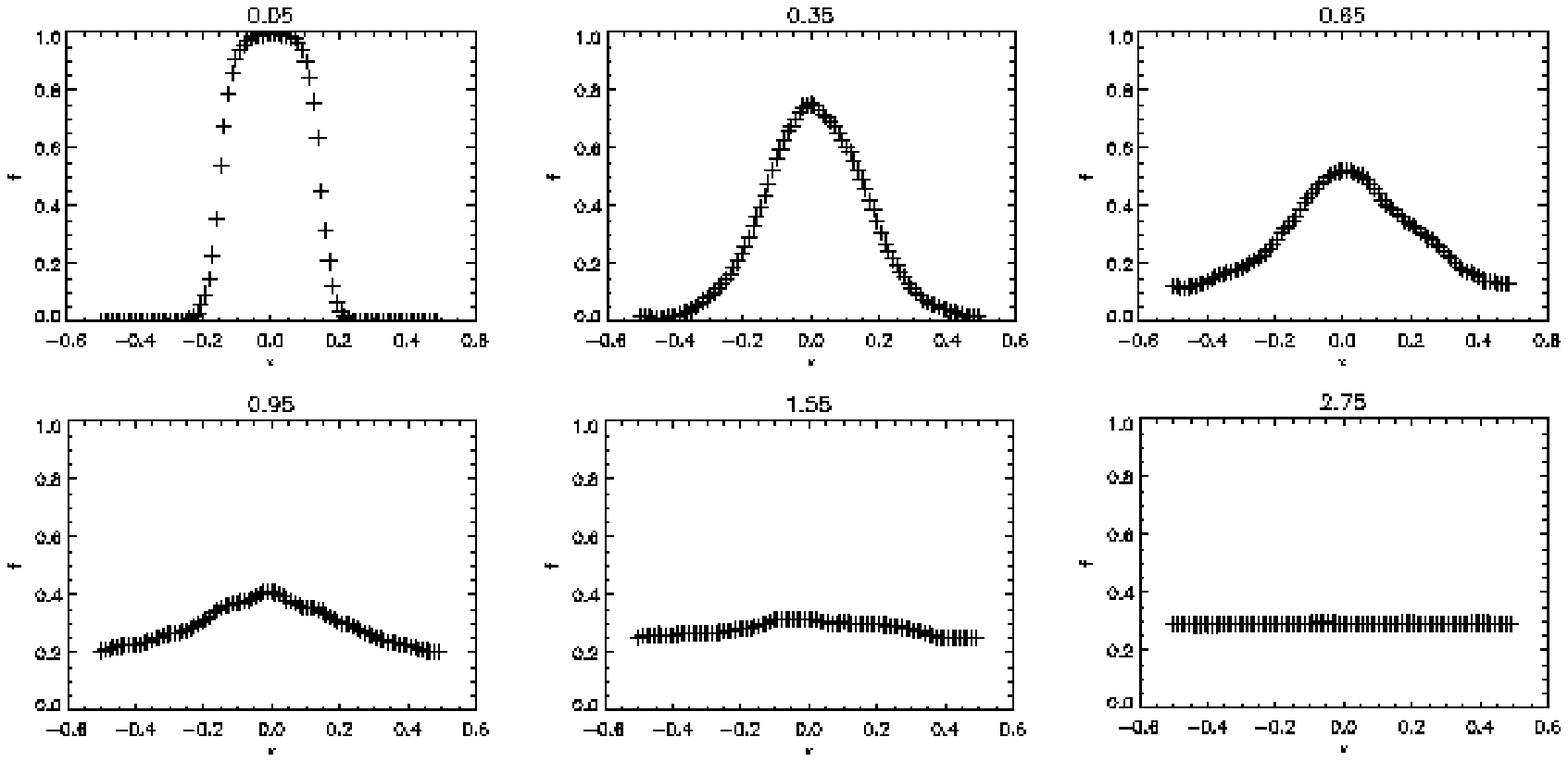}}
\end{picture}
\caption{Evolution of the azimuthally and vertically averaged
radial profile of contaminant fraction. The corresponding time after the
introduction of the contaminant into the turbulent flow (in orbits)
is shown above each frame.  The values of $f$ are shown at each radial grid 
point in the simulation.}
\end{figure*}

In Figure 3 we plot fits of the model curve, equation (8), to
the average radial profiles of contaminant fraction obtained
from the simulations, at the same six
times as shown in Figure 2. 
The value of the diffusion coefficient $D$ that provides the
best fit in each case is shown at the bottom of each panel.  A total of 57 fits
were performed, corresponding to 57 different instants in the evolution of
the contaminant; Figure 3 plots six that are representative of the fitting
process.   The error bars shown correspond to the standard deviation of the simulation data.

\begin{figure*}
\begin{picture}(505,361)
\put(0,0){\includegraphics{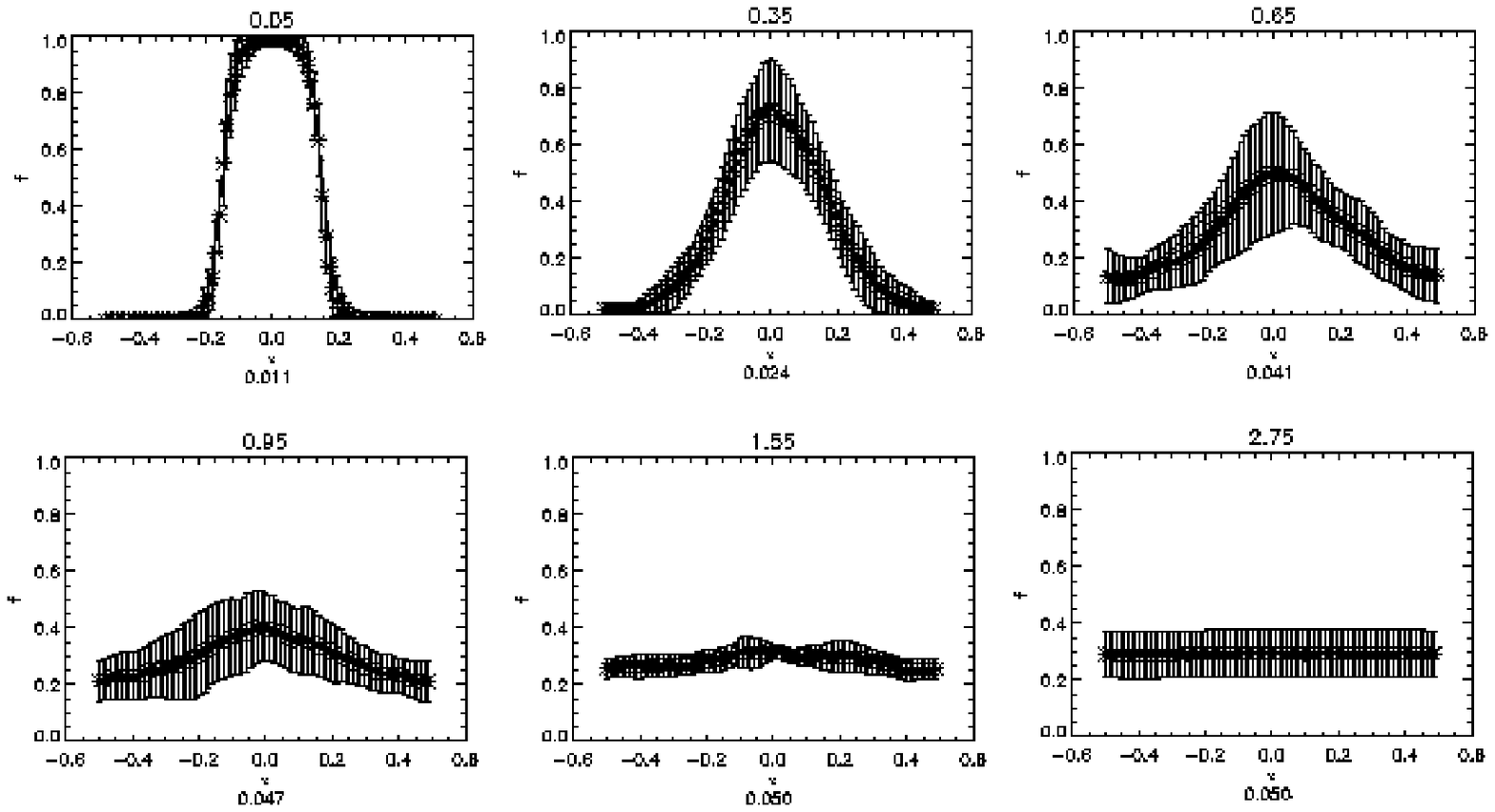}}
\end{picture}
\caption{Fits of the theoretical curves [equation (8)] to the numerically
measured average radial profiles of the contaminant shown in Figure 2. The value of the diffusion coefficient that provides the best fit is shown at the bottom of each panel, in units of $H^2/{\rm orbit}$.}
\end{figure*}

The values of $D$ obtained by fitting equation (8) every 0.05
orbits of evolution are plotted as a function of the time after the
contaminant is introduced in the turbulent flow in Figure 4. During the
first $\sim$ 0.75 orbits of the evolution, $D$ grows from approximately
$1.2 \times 10^{-2} H^2/{\rm orbit}$ to about $5 \times 10^{-2}
H^2/{\rm orbit}$. After about 0.75 orbits, $D$ becomes more nearly
constant; its value does not fall below $\approx 4 \times 10^{-2}H^2/{\rm
orbit}$, and does not exceed $\approx 5 \times 10^{-2}H^2/{\rm orbit}$,
with an average value between 1 and 3 orbits of $(4.7 \pm 1.0) \times
10^{-2}H^2/{\rm orbit}$.  As discussed in the next section, the change in
behavior at 0.75 orbits appears to be related to the spatial extent of
the contaminant, and the size of the largest turbulent eddies in the box.
The diffusion time $t_{\rm diff} \equiv H^2/D$, using the saturated
value of $D$ at late times, is 21 orbits.

\begin{figure*}
\begin{picture}(505,361)
\put(0,0){\includegraphics{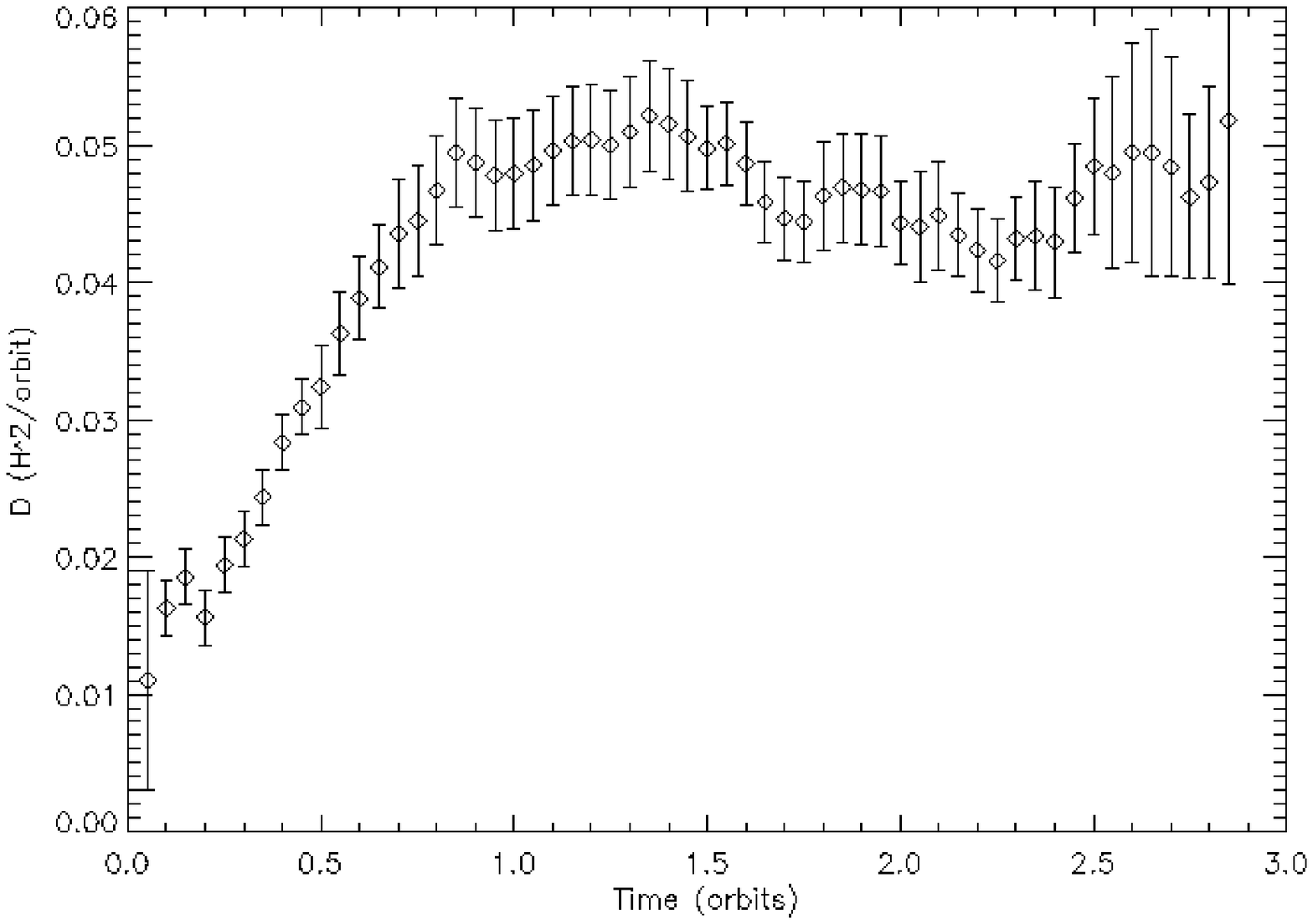}}
\end{picture}
\caption{Turbulent diffusion coefficient $D$,
obtained from the curve-fitting procedure described in the text,
as a function of time.}
\end{figure*}

It is useful to compare the magnitude of $D$ measured from the
simulations with the effective viscosity associated with shear stress,
using the Prandtl number, $P_{\rm eff}\equiv \nu/D$.  The
value of $\nu$ can be calculated through the relation

\begin{equation} \label{eq:nu}
\nu=\alpha c_{\textrm{s}} H=\frac{<<w_{xy}>>}{P}H^2 \Omega
\end{equation}

\noindent where $c_{\textrm{s}}$ is the sound speed,
$<<w_{xy}>>$ is the time- and volume-averaged value of the
$r$-$\varphi$ component of the stress tensor
(magnetic plus kinetic), $H$ is the disc scale height, and $\Omega$
is the angular frequency. The second equality follows from the fact that

\begin{equation}
\alpha=\frac{<<w_{xy}>>}{P}
\end{equation}

\noindent and

\begin{equation}
c_{\textrm{s}}=H \Omega
\end{equation}

Using a value for $<<w_{xy}>>$ which is measured directly from the simulations
(including both the Reynolds and Maxwell stress),
we obtain $\nu \approx 0.52 H^2/{\rm orbit}$. The
Prandtl number is then

\begin{eqnarray}
P_{\rm eff} & = & \frac{\nu}{D} \\
& = & \frac{0.52}{0.047} \\
& \approx & 11
\end{eqnarray}

\section{DISCUSSION}

Using numerical MHD simulations of turbulence driven by the MRI, we have
measured the diffusion coefficient for a passive contaminant mixed by the
flow.  When compared to the effective viscosity associated with angular
momentum transport, the diffusion coefficient is small, that is the
Prandtl number $P_{\rm eff} = \nu/D > 1$.

Of course, numerical errors are an important contribution to the
diffusion rate on scales comparable to the grid spacing.  However, on
larger (resolved) scales up to the box size, there is strong evidence
that numerical effects do not determine the diffusion rate.  The only
numerical errors in solving equation (6) arise in the
advection step of the ZEUS code.  The magnitude and convergence rate
of these errors have been studied in Stone \& Norman (1992a).  For the
second-order van Leer scheme used in these calculations, discontinuities
are diffused until they span 10-15 zones in a few dynamical times,
thereafter very little numerical diffusion occurs.  Figure 2 shows
the profiles reach nearly uniform profiles in less than 3 orbits,
indicative of a much larger diffusion rate than that associated with
numerical errors revealed in the advection tests in Stone \& Norman
(1992a).  Thus, while numerical diffusion may be important on small
scales and at early times, it is unlikely to determine the evolution
observed in Figure 2.  Normally the effects of numerical diffusion would
be assesed by repeating the simulations at different resolutions and
looking for convergence.  However, increasing the numerical resolution
can also change the properties of the turbulence driven by the MRI on
small scales; such resolution studies are less instructive in this case.

Figure 4 shows that the best fit value of the diffusion coefficient
$D$ found by fitting the analytic solution, equation (8),
to the numerically computed profiles increases nearly linearly in time
during the first 0.75 orbits of evolution, and thereafter saturates
at a nearly constant value approximately four times the initial value.
Thus, the turbulent diffusion coefficient seems to acquire a meaningful,
approximately constant value only after the tracer has been spread
over large scales, comparable to the box size.  The possibility that
the diffusion of contaminants in turbulent flows follows a non-Fickian
diffusion law has been suggested previously by Brandenburg et al. (2004).
In our case, the variation of $D$ with time may be related to a variation
in the turbulent velocity fluctuations with scale.  Because turbulence
driven by the MRI follows a nearly Kolmogorov scaling (HGB), the largest
velocity fluctuations are on the largest scales.  As the profile of the
contaminant spreads, it samples larger scales, and larger amplitude
eddies.  Once the profile is spread across the largest scale in the
simulation (the box size), no larger amplitude eddies are possible, and
$D$ becomes approximately constant. At this stage we have not performed
simulations with different box sizes; the effect of changing the box size
may alter the value of $D$, since the amplitude of the MRI turbulence
is set by the box size. However, the stress would also be modified, and
the overall effect would be best examined via their ratio, $P_{\rm eff}$.
  
\indent  The fact that $D$ varies in time for
$t<0.75$ orbits calls into question the applicability of using equation
(7) to describe the evolution of the contaminant at early
times.  A more accurate procedure for measuring $D$ would be to use the
spatial distribution of $f$ at $t=0.75$ orbits as the initial condition
for equation (7), and solve for the resulting evolution, so that
a solution is found only for times when $D$ is known to be constant.
Although this may result in a slightly modified value for $D$ measured
from the simulations, our primary result that $P_{\rm eff} = \nu/D >
1$ will not change. One should also be aware of the possibility that this
transport process follows anomalous diffusion (Bouchaud \& Georges 1990).

If the magnetic field acted as a passive scalar in the flow, and one
could associate the diffusion coefficient $D$ with the coefficient of
resistivity $\eta$, the magnetic Prandtl number would be $P_{\rm m} = \nu/\eta \approx
10$, where $\nu$ is the time- and volume-averaged shear stress (Reynolds
plus Maxwell) in the simulation.  In studies of driven turbulence, Yousef
et al. (2003) found that, for weak magnetic fields, $P_{m}$ is close
to unity.  There are a number of reasons for this discrepancy.  Firstly,
the magnetic field in shearing box simulations of the MRI is clearly
{\em not} a passive scalar, thus $D$ may be a poor proxy for the resistivity.
Perhaps just as important, the momentum transport rate in the shearing
box leads to a much larger effective shear viscosity than in driven
turbulence simulations, because the contributions from the Maxwell stress
are larger than the Reynolds stress (HGB).  Through magnetic torques, momentum
transport can occur indepent of mass motion that would lead to
diffusion of passive contaminants.

The fact that $P_{\rm eff} >> 1$, so that angular momentum transport
is much more efficient than diffusion of contaminants, has important
consequences for the local radial transport of small particles embedded
in a magnetised disc.  Global simulations will be required to determine
the structure and amplitude of radial accretion flows driven by the MRI,
and therefore assess whether small particles can diffusive radially
outwards in the disc.  One of the most important questions for global
simulations of accretion discs is to determine whether large scale
poloidal magnetic fields can be advected inwards by an accretion flow
(leading to strong vertical fields in the inner regions of the disc),
or whether they can diffuse outwards more rapidly than they are carried
in (Clarke \& Pringle 1988).  If the diffusion coefficient $D$ measured
here were representative of the vertically averaged resistive diffusion
coefficient for poloidal magnetic fields, then $P_{m} >> 1$ would
indicate fields would be advected inward.  However, since poloidal
magnetic fields drive the MRI, and since reconnection (which is an
important saturation mechanism in the nonlinear regime of the MRI) can
break flux-freezing, it is unlikely equation (6) can be used
to describe the radial evolution of poloidal fields.  Global simulations
of thin Keplerian discs which can address these issues are needed.

\bigskip

{\bf Acknowledgements:}
AC acknowledges support from CONACYT scholarship 167912, thanks the Department of Astrophysical Sciences at Princeton University for their hospitality, and would like to thank Gordon Ogilvie for fruitful discussions.
JS is grateful for financial support from the Royal
Society, the University of Cambridge, and NSF grant AST-0413788
for portions of this work.

\end{document}